\documentclass[aps,prl,reprint,twocolumn,floatfix,preprintnumbers,nofootinbib,superscriptaddress]{revtex4-2}
\usepackage{graphicx,color,xcolor}
\usepackage{amsmath,amssymb,amsfonts,amsthm}
\usepackage{wasysym}
\usepackage[caption=false]{subfig}
\usepackage[colorlinks=True, citecolor=blue, urlcolor=blue, linkcolor=blue]{hyperref}
\usepackage{ulem}
\usepackage{comment}
\newcommand{\be}{\begin{equation}}
\newcommand{\ee}{\end{equation}}
\newcommand{\ba}{\begin{eqnarray}}
\newcommand{\ea}{\end{eqnarray}}

\newcommand{\jmax}{j_\mathrm{max}}

\newcommand{\bn}{{\boldsymbol n}}
\newcommand{\cF}{{\mathcal F}}

\begin{document}

\title{The Magic Barrier before Thermalization}
\author{Lukas Ebner}
\affiliation{Max Planck Institute of Quantum Optics, 85748 Garching, Germany}
\affiliation{Munich Center for Quantum Science and Technology (MCQST), 80799 Munich, Germany}
\author{Berndt M\"uller}
\affiliation{Department of Physics, Duke University, Durham, North Carolina 27708, USA}
\author{Andreas Sch\"afer}
\affiliation{Institut f\"ur Theoretische Physik, Universit\"at Regensburg, D-93040 Regensburg, Germany}
\affiliation{Department of Physics, National Taiwan University, Taipei, Taiwan 106}
\author{Leonhard Schmotzer}
\affiliation{Institut f\"ur Theoretische Physik, Universit\"at Regensburg, D-93040 Regensburg, Germany}
\author{Clemens Seidl}
\thanks{\href{mailto:clemens.seidl@physik.uni-regensburg.de}{clemens.seidl@physik.uni-regensburg.de}; corresponding author and leading contributor to the results in this work.}
\affiliation{Institut f\"ur Theoretische Physik, Universit\"at Regensburg, D-93040 Regensburg, Germany}
\affiliation{Department of Physics and Arnold Sommerfeld Center for Theoretical Physics (ASC), Ludwig Maximilian University of Munich, 80333 Munich, Germany}
\affiliation{Munich Center for Quantum Science and Technology (MCQST), 80799 Munich, Germany}
\author{Xiaojun Yao}
\affiliation{InQubator for Quantum Simulation, Department of Physics,
University of Washington, Seattle, Washington 98195, USA}
\date{\today}
\preprint{IQuS@UW-21-110}

\begin{abstract}
We investigate the time dependence of anti-flatness in the entanglement spectrum, a measure for non-stabilizerness and lower bound for non-local quantum magic resource, on a subsystem of a linear SU(2) plaquette chain during thermalization. Tracing the time evolution of a large number of initial states, we find that the anti-flatness exhibits a barrier-like maximum during the time period when the entanglement entropy of the subsystem grows rapidly from the initial value to the microcanonical entropy. The location of the peak is strongly correlated with the time when the entanglement exhibits the strongest growth. This behavior is found for generic highly excited initial computational basis states and persists for coupling constants across the ergodic regime, revealing a universal structure of the entanglement spectrum during thermalization. We conclude that quantitative simulations of thermalization for nonabelian gauge theories require quantum computing. We speculate that this property generalizes to other quantum chaotic systems, a conjecture supported by analogous behavior observed in real-time simulations of the mixed-field Ising model.
\end{abstract}

\maketitle

{\it Introduction.} Quantum computers are expected to eventually outperform classical supercomputers for a number of important applications by harvesting the resources opened up by quantum entanglement of qubits or qudits. However, identifying such applications is not as straightforward as one might imagine. While it is easy to prepare highly entangled quantum states, in particular by the Clifford set of gates comprising the Hadamard gate, the phase gate and the two-qubit CNOT gate, this entanglement alone does not enable dramatic gains in computational power. In fact, the Gottesman-Knill (GK) theorem \cite{Gottesman:1998hu} states that any quantum circuit built solely from Clifford gates (also known as the stabilizer circuits) can be simulated in polynomial time on a classical digital computer. 

Not all unitary quantum circuits are stabilizer circuits. A universal gate set requires at least one additional gate, e.g., the $\pi/8$ (T) gate. The GK theorem implies that only quantum circuits requiring such non-Clifford gates can realize true quantum advantage. 

In practice, it would be most helpful to have a quantitative measure of the potential advantage of quantum computation over classical computation for simulating a given quantum system. One such measure has become known as non-stabilizerness or ``magic'' \cite{Bravyi:2004isx,Aaronson:2004xuh,Bravyi:2012lxn,Emerson:2013zse}, but its experimental or numerical determination is challenging. Recently, the Stabilizer R\'enyi Entropy (SRE) was introduced as a computable measure of non-stabilizerness \cite{Leone:2021rzd} and has been studied in many systems including matrix product states \cite{Haug:2022vpg}, random quantum circuits \cite{Turkeshi:2024pnj} and  states \cite{Turkeshi:2023lqu}, lattice models and gauge theories \cite{Tarabunga:2023ggd,Odavic:2024lqq,Santra:2025dsm,Falcao:2024msg,Tirrito:2024kts}, nuclei \cite{Brokemeier:2024lhq}, neutrinos \cite{Chernyshev:2024pqy} and Bell inequalities~\cite{Cusumano:2025zbx}. 
In high-energy proton-proton reaction at the Large Hadron Collider (LHC), identifying quantum systems with non-stabilizerness involves studying certain processes such as the top quark-antiquark pair ($t\bar{t}$) production \cite{ATLAS:2023fsd,Barr:2024djo,CMS:2024pts,White:2024nuc,CMS:2024zkc}. In part, the investigation of entanglement at LHC arises from their promise to provide a novel avenue in search for possible signals of physics beyond the Standard Model. For instance, the authors of \cite{CMS:2025cim} considered a general form of non-stabilizerness
for mixed quantum states, which is defined based on the second SRE \cite{Leone:2021rzd,White:2024nuc}. They found the result of their analysis for the $t\bar{t}$ data to agree with the Standard Model expectations.

A quantity closely related to non-stabilizerness is the anti-flatness of the entanglement spectrum of a quantum system \cite{Tirrito:2023fnw}. The reduced density matrix $\rho_A$ of a subsystem $A$ is defined by
\begin{equation}
\rho_A= {\rm Tr}_{A^c}(\rho) \,,
\end{equation}
where $A^c$ denotes the complement of the subsystem. The entanglement entropy 
\begin{equation}
S_A = -{\rm Tr}(\rho_A\log\rho_A) \,,
\end{equation} 
provides a measure of the entanglement in the full system wave function. Li and Haldane \cite{Li:2008kda} showed that the spectrum of the so-called entanglement Hamiltonian $H_A = -\log\rho_A$ contains additional information beyond just the total amount of entanglement, which has been used to study thermalization in $Z_2$ lattice gauge theory \cite{Mueller:2021gxd}. If this spectrum is flat, the so-called anti-flatness 
\begin{equation}
\cF_A = {\rm Tr}(\rho_A^3) - [{\rm Tr}(\rho_A^2)]^2 \,,
\end{equation}
vanishes. Refer to \cite{Turkeshi:2023ctq} for another related measure.

It has been shown that non-stabilizerness can be decomposed into local and non-local contributions \cite{Cao:2024nrx, Cao:2023mzo} and $\cF_A$ provides a lower bound on non-local non-stabilizerness~\cite{Cao:2024nrx}.
While local non-stabilizerness characterizes non-Clifford features that can be attributed to individual subsystems and thus removed by local unitaries, non-local contribution captures intrinsically multipartite, non-stabilizer correlations that cannot be eliminated locally and require entanglement to exist \cite{Cao:2024nrx, Cao:2023mzo, Qian:2025oit, Andreadakis:2025mfw}.
The total amount of non-stabilizerness can be viewed as the sum of these two components, with the non-local one quantifying genuinely non-classical correlations beyond both stabilizer structure and entanglement.
Hence, anti-flatness bounds the hardness of classical simulations from below.

Here we pursue the question whether the description of thermalization of an isolated quantum system requires quantum computing, using SU(2) lattice gauge theory on a linear plaquette chain as an example. This system can be viewed as an extreme simplification of the more complex process of thermalization in a highly excited system of quarks and gluons as it is created in relativistic heavy ion collisions \cite{Busza:2018rrf}. We will show that in this model system the time dependence of anti-flatness of the entanglement spectrum and the growth of the entanglement entropy are closely connected. While not constituting a general proof, our results support the notion that an accurate description of gauge field thermalization requires quantum computing.

{\it Method.} We investigate pure SU(2) lattice gauge theory (LGT) in $2+1$ dimensions as a representative framework for exploring generic nonabelian gauge dynamics. To access real-time phenomena, we employ the Hamiltonian formulation of SU(2) LGT, which provides a natural setting for studying out-of-equilibrium processes beyond the limitations of Euclidean lattice and perturbative approaches. The discretized Kogut-Susskind (KS) Hamiltonian for unit lattice spacing can be written as \cite{Kogut:1974ag}
\begin{align}
    H=\frac{g^2}{2}\sum_{\rm links}(E^a_i)^2-\frac{2}{g^2}\sum_{\bn}\Box(\bn) \,,
\label{eq:KS_Hamiltonian}
\end{align} 
where $g$ is the coupling constant, $E^a_i$ is the electric field operator along the direction $i=\hat{x}$ or $\hat{y}$ with the SU(2) index $a$ (both of which are implicitly summed over), and $\Box(\bn)$ denotes the plaquette operator at $\bn = (n_x, n_y)$, i.e., the trace of the product of four link variables (Wilson lines) around an elementary plaquette.
The first and second part of the Hamiltonian represent the electric and magnetic energy of the gauge field respectively.

The discretized KS Hamiltonian can be represented in the electric basis, which labels the state on each link by the SU(2) quantum numbers $|jm_Lm_R\rangle$. In this basis the electric energy $(E_i^a)^2$ is diagonal with eigenvalues $j(j+1)$. The matrix elements of the plaquette operator may be expressed as a combination of Wigner 6-$j$ symbols \cite{Klco:2019evd}. Truncation of the local Hilbert space of each link to representations with $0 \leq j \leq \jmax$ renders the total Hilbert space finite-dimensional and allows for exact numerical diagonalization of the Hamiltonian.

In this Letter we study an aperiodic seven-plaquette chain with the electric field representation truncated at $\jmax=1$, but our investigation can also be performed for other electric field truncations and boundary conditions.
\begin{figure}
    \centering
    \includegraphics[width=\linewidth]{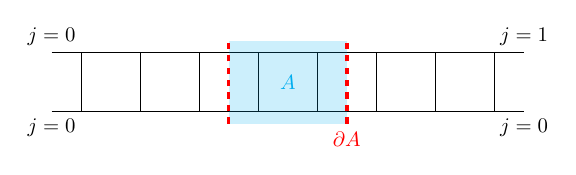}
    \caption{Seven-plaquette chain with asymmetric boundary conditions. Subsystem $A$ is defined as the middle plaquette and adjacent links.}
    \label{fig:fig1_plaquette_chain}
\end{figure}
Fig.~\ref{fig:fig1_plaquette_chain} schematically illustrates the corresponding link configuration of the system.
In the specific case where vertices contain at most three adjoining links, as here, the physical states of the theory are unambiguously represented by the irreducible representation $j$ of each link variable~\cite{Klco:2019evd,ARahman:2022tkr}. Asymmetric boundary conditions are enforced by setting the external links to $\{j_{\rm ext}\}=\{0,0,0,1\}$, such that the Hamiltonian has no spatial symmetries\footnote{Otherwise, the Hilbert space would decompose into symmetry sectors, whose basis states fall into different symmetry classes. The analysis can be done analogously, but basis states within a symmetry sector may have different initial values of entanglement entropy.}.
The Hilbert space dimension of the seven-plaquette SU(2) chain with $\jmax=1$ and the above boundary conditions is 36334.
Similar systems were studied in \cite{Yao:2023pht,Ebner:2023ixq,Ebner:2024mee,Ebner:2024qtu}, where it was demonstrated that these quantum systems exhibit both ergodic and non-ergodic coupling regimes. In particular, for sufficiently large $g^2$, the magnetic contribution to the Hamiltonian becomes too small to generate chaotic energy-level statistics on a small lattice. The KS Hamiltonian has two integrable limits at a given lattice size: $g^2\rightarrow\infty$ and $g^2\rightarrow0$, but for intermediate (ergodic) coupling (here, $0.4\lesssim g^2\lesssim 1.5$)\footnote{For the system under investigation we find the mean restricted gap ratio of the energy level spectrum to be $\{\langle r\rangle\}\approx\{0.496, 0.519, 0.531, 0.530, 0.530, 0.527, 0.520, 0.505\}$ for $\{g^2\}=\{0.3, 0.4, 0.6, 0.8, 1.0, 1.2, 1.5, 1.7\}$, compared to $\langle r\rangle_{\rm GOE}\approx0.531$~\cite{Atas:2013gap}.}, the eigenvalue distribution of the seven-plaquette system follows closely the prediction for a Gaussian orthogonal ensemble (GOE) and the system exhibits quantum chaos.
In this Letter, we focus specifically on the ergodic coupling regime of the SU(2) LGT on seven plaquettes, motivated by the fact that the classical SU(2) theory is chaotic~\cite{Bolte:1999th} and, upon quantization, is expected to obey the eigenstate thermalization hypothesis~\cite{Yao:2023pht,Ebner:2023ixq,Das:2025utp}.

To compute entanglement entropy and anti-flatness for subsystems of the plaquette chain we use the technique described in \cite{Ebner:2024mee}. Each electric basis state can be written as a tensor product of the $j$-quantum numbers of individual links $|\{j\}\rangle$, subject to the Gauss law constraint for physical states. The boundary of a subsystem $A$ cuts through its adjacent horizontal links such that the reduced density matrix remains invariant under time-independent gauge transformations at each vertex (see Fig.~\ref{fig:fig1_plaquette_chain}). The cut links result in a direct sum structure in the reduced density matrix $\rho_A$, which can be diagonalized to obtain the full entanglement spectrum, the entanglement entropy $S_A$ and the anti-flatness $\cF_A$.
The subsystem $A$ considered in the following consists of the middle plaquette with adjacent cut links as depicted in Fig.~\ref{fig:fig1_plaquette_chain}.

To analyze the thermalization dynamics, we initialize the system in pure states that are not eigenstates of the Hamiltonian but whose mean energy falls into a given energy window. To achieve this, we employ computational basis states (electric energy eigenstates) within the desired energy window.
Numerical calculations were performed using our custom code available at \cite{ebner_2025_16739090}.

{\it Time dependence of entanglement and anti-flatness.}
\begin{figure*}[th]
    \centering
    \includegraphics[]{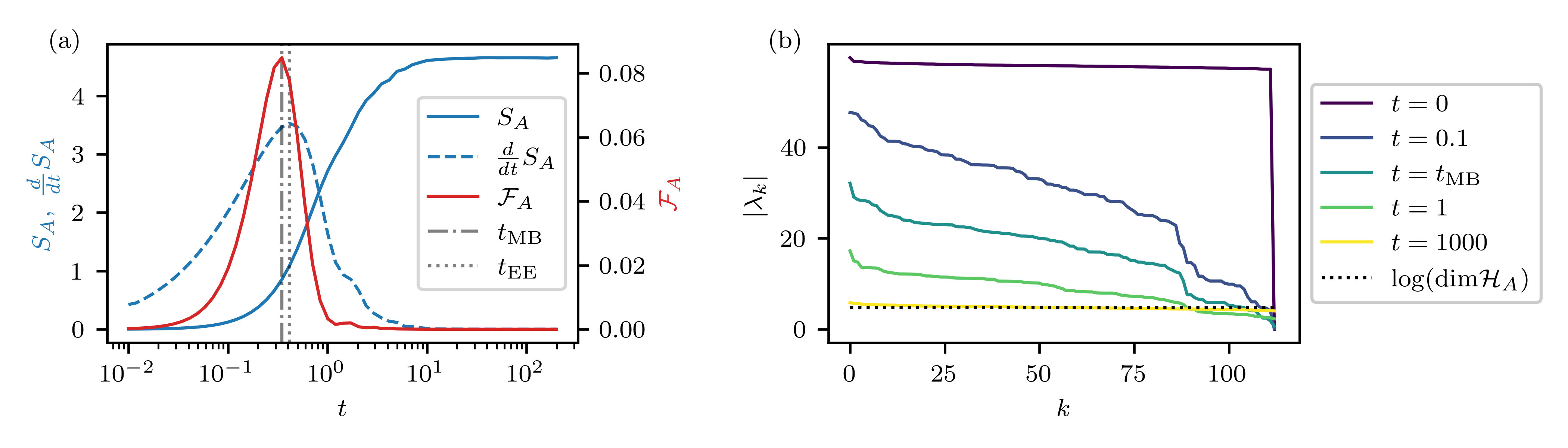}
    \caption{(a) Real-time evolution of entanglement entropy $S_A$ (blue solid line, left scale) and anti-flatness $\mathcal{F}_A$ (red solid line, right scale) of a small one-plaquette subsystem with cut links in the middle of an aperiodic seven-plaquette chain with $\jmax=1$, ergodic coupling $g^2=1$ and asymmetric boundary conditions $\{j_{\rm ext}\}=\{0,0,0,1\}$ for a randomly chosen, highly excited initial electric basis state with energy $E-E_0\approx 19.17$, where $E_0$ is the ground state energy. The blue dashed line shows the entanglement entropy growth rate. $t_{\rm MB}$ and $t_{\rm EE}$, indicated by gray lines, correspond to the time of the magic barrier and maximum entanglement growth rate, respectively.
    (b) Entanglement spectra of the above state at different times, including $t_{\rm MB}\approx0.348$, during the thermalization process. 
    $\lambda_k$ denotes the $k$-th eigenvalue of the entanglement Hamiltonian $H_A=-\log\rho_A$. The flat-spectrum limit is given by $|\lambda_k|= \log({\rm dim}(\mathcal{H}_A)) \approx 4.727$.
      }
    \label{fig:EEvsAF_single_state}
\end{figure*}
In order to understand the dynamics of quantum complexity during thermalization, we track the time dependence of entanglement entropy $S_A(t)$ and anti-flatness $\cF_A(t)$ of the subsystem $A$ for the chosen globally evolving pure states. For $S_A$ this evolution is well understood. Starting from an initial pure product state, we have $S_A(0)=0$. As the unitary dynamics proceeds, entanglement spreads and the entanglement entropy rises steadily until it saturates at a value consistent with the thermal entropy of the subsystem \cite{Ebner:2024qtu}. This behavior, shown by the solid blue line in Fig.~\ref{fig:EEvsAF_single_state}~(a) for an arbitrarily chosen highly excited initial electric basis state, reflects the effective thermalization of the subregion: while the global state remains pure, the local reduced state approaches a mixed quasi-thermal ensemble, as expected for a system obeying the eigenstate thermalization hypothesis \cite{Deutsch:1991msp, Srednicki:1994mfb, DAlessio:2015qtq}.

In contrast, the anti-flatness $\cF_A(t)$, being a witness of non-stabilizerness, displays a more intricate profile, as shown by the solid red line in Fig.~\ref{fig:EEvsAF_single_state}~(a). At early time, the subsystem has negligible non-local non-stabilizerness since it is close to a classical configuration. As the dynamics scrambles information across the full system, the reduced state $\rho_A$ becomes highly non-classical, resulting in a sharp rise of the anti-flatness.
As $S_A(t)$ approaches saturation, $\cF_A(t)$ falls again steeply as $\rho_A$ approaches a nearly maximally mixed form characteristic of a high-energy system in (microcanonical) thermal equilibrium, which can be efficiently studied on classical digital computers. The peak of the anti-flatness appears at time $t_{\rm MB}$, which approximately equals the time $t_{\rm EE}$ when $S_A(t)$ is growing most rapidly.

The dynamics of the entanglement spectrum can also be visualized directly as in Fig.~\ref{fig:EEvsAF_single_state}~(b), where we show the eigenvalues of the entanglement Hamiltonian at various times during the thermalization process. The initially flat spectrum develops a visible non-flat structure during the time when $\cF_A(t)$ is non-vanishing. At late times, in thermal equilibrium, the spectrum returns to an almost flat (uniform) structure where all eigenvalues approximately approach $\log({\rm dim}(\mathcal{H}_A))$, where $\mathcal{H}_A$ denotes the Hilbert space on the subsystem $A$. The appearance of the anti-flatness peak signals that the subsystem acquires strong ``quantumness'' during the thermalization process. However, the quantum features detected by the anti-flatness are transient: they dominate the intermediate time regime of thermalization, but are washed out again as the subsystem relaxes toward thermal equilibrium.
Similar barrier-like behavior was also found for non-stabilizerness in state preparation algorithms such as adiabatic annealing \cite{Capecci:2025ull} and for operator entanglement dynamics in conformal field theory \cite{Wang:2019ued}.

Together, entanglement entropy and anti-flatness offer complementary insights: the former captures the irreversible buildup of correlations between the subsystem and its complement, while the latter identifies the transient temporal regime when the dynamics is the most ``quantum'' and the most difficult to simulate classically.

{\it Universal magic barrier and entropy growth.}
\begin{figure*}[t]
    \centering
    \includegraphics[]{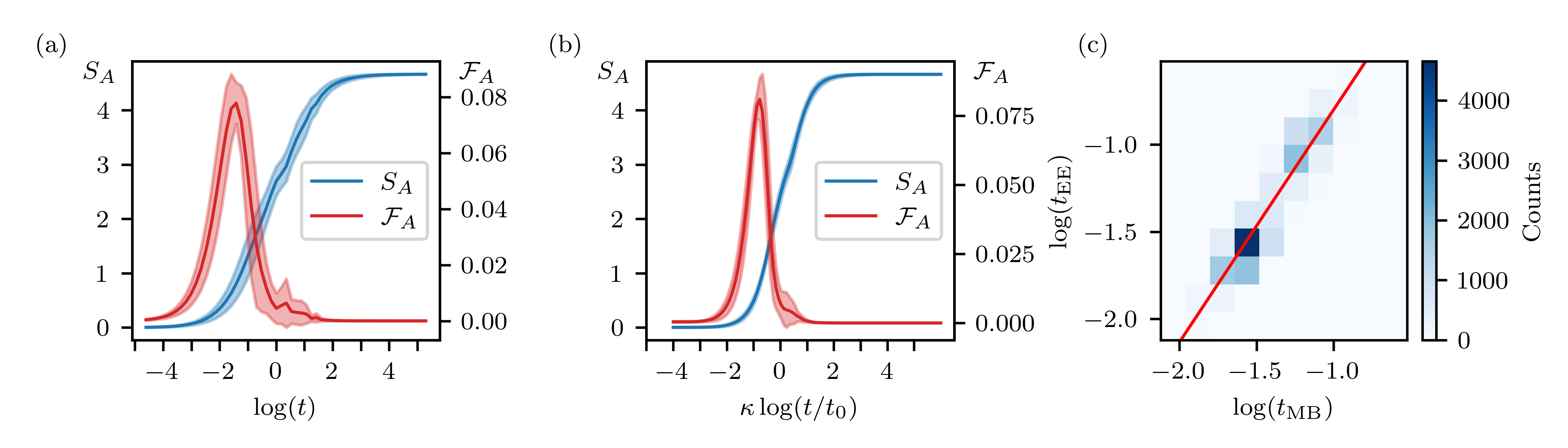}
    \caption{(a) Real-time evolution of $S_A$ and $\mathcal{F}_A$ for a one-plaquette subsystem with cut links in the middle of an aperiodic seven-plaquette chain with $\jmax=1$ and asymmetric boundary conditions $\{j_{\rm ext}\}=\{0,0,0,1\}$ at ergodic coupling $g^2=1$ for all electric basis states within the highly excited energy window $E-E_0\in [19.71, 20.21]$. The energy window contains 2728 states.
    (b) Rescaled time evolution of the same system and state ensemble. The real time $t$ is replaced by $\kappa\log(t/t_0)$ with state-dependent fit parameters $\kappa$ and $t_0$, such that the thermalization of different states is synchronized.
    The solid lines and bands in (a) and (b) are the ensemble means and $1\sigma$-bands. The left and right scales correspond to $S_A$ and $\mathcal{F}_A$, respectively.
    (c) 2D histogram showing the joint distribution of magic barrier time $t_{\rm MB}$ and time of maximum entanglement entropy growth $t_{\rm EE}$ on logarithmic scales for all 18389 physical electric basis states of the above system with energies below the spectrum mean. Color intensity corresponds to bin counts. The red line represents a linear fit with slope 1.331 and intercept 0.530.
    }
    \label{fig:fig3_MBvsEEgrowth}
\end{figure*}
We now study whether the appearance of the magic barrier is state-independent and quantify the correlation between the time of the anti-flatness peak and that of the maximum growth rate of the entanglement entropy. We compute the time evolution of entanglement entropy and anti-flatness of a small subsystem with one plaquette and dangling links in the middle of an aperiodic seven-plaquette chain with $\jmax=1$ and asymmetric boundary conditions $\{j_{\rm ext}\}=\{0,0,0,1\}$ for all 2728 electric basis states inside the highly excited energy window $E-E_0\in[19.71, 20.21]$, where $E_0$ is the ground state energy.

The result is shown in Fig.~\ref{fig:fig3_MBvsEEgrowth}~(a). It can be seen that at early time, $\log(t)<-1$, the entanglement entropies of different initial states grow at different rates. This is analogous to the behavior of classical chaotic systems with a broad Lyapunov spectrum, where the coarse-grained entropies for different initial states are found to grow at different rates at early time, depending on their degree of overlap with the modes characterized by large Lyapunov exponents. However, in all cases the anti-flatness exhibits a broad, nevertheless very prominent, peak during the period when the state's entanglement grows. It is important to emphasize that we analyze all electric basis states in the energy window of interest and observe no outliers, i.~e., all states exhibit the same qualitative behavior.

In order to better analyze the profile of the magic barrier with respect to the thermalization process, we make use of the universal behavior of entanglement entropy growth for subsystems found in \cite{Ebner:2024qtu}.
We fit the corresponding time evolution for each state to the function $S_A(t) = S_A(0)+\frac{S_A(\infty)-S_A(0)}{1+(t/t_0)^{-2\kappa}}$, where $S_A(0)=0$ and $S_A(\infty)$ approaches a quasi-thermal value that only depends on the state's energy. The fit parameters $\kappa$ and $t_0$ correspond to a thermalization speed and time scale, respectively. By using $\kappa\log(t/t_0)$ instead of $t$ as a ``thermalization clock'' we effectively synchronize the thermalization process of each state. As a result, as seen in Fig.~\ref{fig:fig3_MBvsEEgrowth}~(b), the entanglement entropy and anti-flatness as functions of this ``thermalization clock'' show smaller variations and more manifest functional profiles.

\begin{figure*}[th]
    \centering
    \includegraphics[]{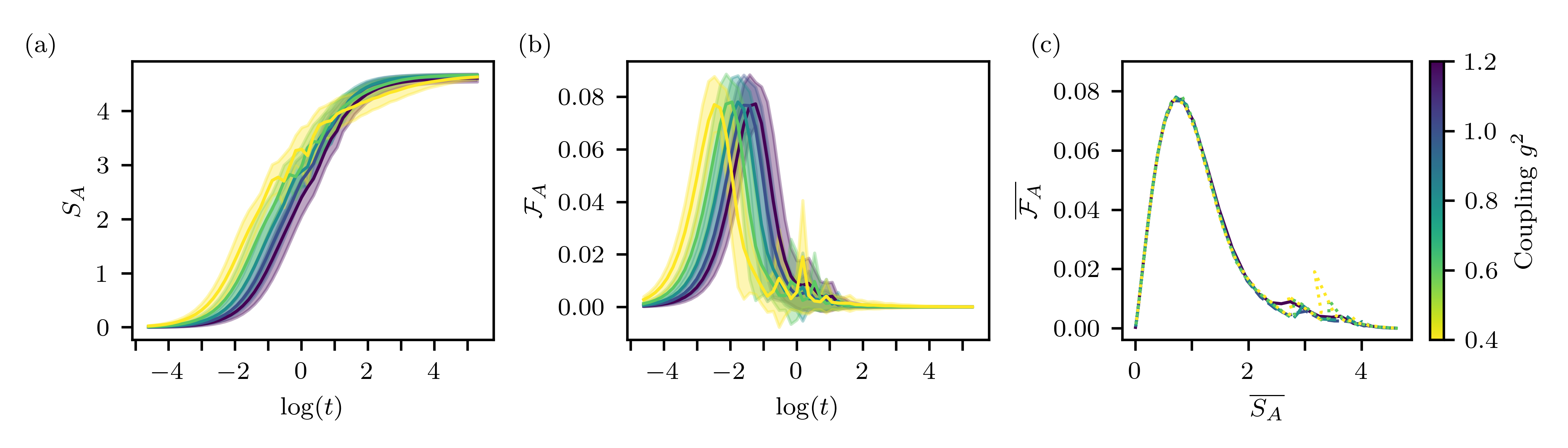}
    \caption{Coupling dependence of entanglement dynamics. Real-time evolution of (a) $S_A$ and (b) $\mathcal{F}_A$ for a small one-plaquette subsystem with cut links in the middle of an aperiodic seven-plaquette chain with $\jmax=1$ and asymmetric boundary conditions $\{j_{\rm ext}\}=\{0,0,0,1\}$ at different ergodic couplings $g^2\in\{1.2, 1.0, 0.8, 0.6, 0.4\}$. The ensemble of electric basis states is chosen such that at $g^2=0.6$ they lie in the highly excited energy window $E-E_0\in [26.48, 26.98]$. This ensemble contains 2995 states. The solid lines and bands in (a) and (b) are the ensemble means and $1\sigma$-bands.
    (c) Relations between the ensemble means $\overline{S}_A$ an $\overline{\cF}_A$ for different ergodic couplings, which exhibit very mild coupling dependence.
    }
    \label{fig:fig4_coupling_dependence}
\end{figure*}

Qualitatively, the functional profiles suggest that the magic barrier emerges around the time when the entanglement entropy grows the fastest. To quantify this observation, we compare the magic barrier time $t_{\rm MB}$ to the time $t_{\rm EE}$ of maximal entanglement entropy growth, for all physical electric basis states—defined as those with energies below the mean of the spectrum. This physical ensemble contains 18389 non-eigenstates, for which a 2D histogram of $\log(t_{\rm MB})$ and $\log(t_{\rm EE})$ are depicted in Fig.~\ref{fig:fig3_MBvsEEgrowth}~(c). 
We find a strong positive correlation between the two quantities. The Pearson correlation coefficient between $\log (t_{\rm MB})$ and $\log (t_{\rm EE})$ is $r=0.927$,  indicating a highly significant relationship.
Furthermore, as indicated by the color distribution of the histogram, the time scales $t_{\rm MB}$ and $t_{\rm EE}$ roughly coincide for the majority of investigated states.

{\it Coupling constant dependence.}
In order to make predictions about the emergent magic barrier for the physical limit of SU(2) LGT, it is necessary to study its coupling dependence. With respect to the continuum limit, we are especially interested in the weak coupling limit, where $g^2$ is small but the theory is still ergodic.

We perform similar analyses as above for four other coupling values $g^2=0.4$, $0.6$, $0.8$ and $1.2$, all of which exhibit quantum chaotic energy level statistics.
The ensemble of initial electric basis states for all five couplings is chosen such that for $g^2=0.6$ they lie in the highly excited, narrow energy window $E-E_0\in [26.48, 26.98]$.
In Fig.~\ref{fig:fig4_coupling_dependence}~(a), we compare the entanglement entropy growth of this ensemble for the different couplings and find that the growth is faster at smaller coupling. In particular, we find a maximum of average growth rate $(d\overline{S}_A/dt)_{\max}=\{3.631, 4.443, 5.665, 7.563, 11.106\}$ for $g^2=\{1.2, 1.0, 0.8, 0.6, 0.4\}$, consistent with a $(d\overline{S}_A/dt)_{\max} \propto g^{-2}$ scaling, which reflects the $g^{-2}$ factor controlling the magnetic energy (plaquette operator) of the KS Hamiltonian \eqref{eq:KS_Hamiltonian} that controls the mixing rate between electric basis states. As a result, the computational basis states, which are eigenstates of the electric energy, get scrambled faster under unitary time evolution when the strength of the coupling is reduced.

The time $t_{\mathrm{EE}}$ at which the entanglement entropy growth reaches its maximum rate lies within a linear growth regime, consistent with the findings in \cite{Toniolo:2024tal}. 
The quantity $(dS_A/dt)_{\max}$ corresponds to the slope in this linear regime and can be interpreted as the entanglement velocity, i.e., the characteristic speed of the wave-front carrying entanglement from $A^c$ into $A$ \cite{Liu:2013iza}.

Furthermore, these findings are in accordance with the time evolution of anti-flatness. In Fig.~\ref{fig:fig4_coupling_dependence}~(b), the magic barrier is seen to emerge earlier for decreasing coupling, while maintaining its height and shape.
Instead of tracing entanglement entropy and anti-flatness as functions of time individually, we can also calculate $\cF_A$ as a function $S_A$ and thus obtain a thermalization profile of the magic barrier instead of a temporal profile. This profile is found to be almost identical for all tested ergodic couplings, as shown in Fig.~\ref{fig:fig4_coupling_dependence}~(c), where both quantities are ensemble-averaged.
These results show that the entanglement spectrum has a coupling-independent structure and the magic barrier emerges during the equilibration process, as long as the system is ergodic and the subsystem thermalizes.
In \cite{sm}, we show the coupling-independent structure of the entanglement spectrum exemplarily for three randomly chosen states at $t=t_{\rm MB}$.

{\it Other quantum chaotic models.}
In order to investigate whether the magic barrier phenomenon is special to nonabelian gauge theories or may also be observed in other quantum ergodic models, we repeat the main analysis of this work for the mixed-field Ising model. The results shown in \cite{sm} indicate that the anti-flatness of a small subsystem exhibits a similar barrier-like maximum during the time when its entanglement entropy grows rapidly from the initial value to the microcanonical entropy. Further, for the most ergodic field strengths, we find a coupling-independent relation between the anti-flatness and entanglement entropy during local equilibration, similar to the one found for the SU(2) gauge theory.

{\it Conclusion.}
We have studied the equilibration dynamics of subsystem entanglement entropy and anti-flatness under unitary evolution in a (2+1)-dimensional SU(2) lattice gauge theory constrained on a linear plaquette chain with $\jmax=1$. We find that the anti-flatness, a lower bound on non-local non-stabilizerness, exhibits a universal barrier-like peak during the period of fast entanglement growth. This behavior occurs for all highly excited electric basis states and persists across ergodic couplings, revealing a coupling-independent structure of the entanglement spectrum during thermalization. The timing of the barrier is strongly correlated with that of the maximal entanglement growth, demonstrating that the thermalization dynamics in the SU(2) gauge theory involves simultaneous generation of high non-stabilizerness and high entanglement.
Supported by similar behavior found for the mixed-field Ising model, we
speculate that this generic barrier phenomenon may persist across a broad class of quantum chaotic theories, implying that quantitative simulations of thermalization in these theories inherently require quantum computation.

Promising directions for future work include testing this conjecture in other chaotic systems—most notably SU(3) gauge theory, examining the evolution of anti-flatness during specific physical processes such as string breaking in fermionic models and glueball formation~\cite{Grieninger:2026bdq,Cataldi:2025cyo,Cao:2026qky}, and computing non-stabilizerness directly via measures for qudit systems rather than through its anti-flatness bound. We plan to pursue these investigations in future work.

\begin{acknowledgments}
{\it Acknowledgments.}
The authors gratefully acknowledge the scientific support and HPC resources provided by the Erlangen National High Performance Computing Center (NHR@FAU) of the Friedrich-Alexander-Universität Erlangen-Nürnberg (FAU).
B.M. acknowledges support by the U.S. Department of Energy, Office of Science (Grant DE-FG02-05ER41367) and by the National Science Foundation (Project PHY-2434506).
X.Y. is supported by the U.S. Department of Energy, Office of Science, Office of Nuclear Physics, InQubator for Quantum Simulation (IQuS) (https://iqus.uw.edu) under Award Number DOE (NP) Award DE-SC0020970 via the program on Quantum Horizons: QIS Research and Innovation for Nuclear Science and acknowledges the discussions at the ``Many-Body Quantum Magic" workshop held at the IQuS hosted by the Institute for Nuclear Theory in Spring 2025.
L.E.~acknowledges funding by the Max Planck Society, the Deutsche Forschungsgemeinschaft (DFG, German Research Foundation) under Germany’s Excellence Strategy – EXC-2111 – 390814868, and the European Research Council (ERC) under the European Union’s Horizon Europe research and innovation program (Grant Agreement No.~101165667)—ERC Starting Grant QuSiGauge. Views and opinions expressed are those of the author(s) only and do not necessarily reflect those of the European Union or the European Research Council Executive Agency. Neither the European Union nor the granting authority can be held responsible for them. This work is part of the Quantum Computing for High-Energy Physics (QC4HEP) working group.
A.S. and C.S. are supported by the DFG (German Research Foundation, grant 553079183).
C.S. acknowledges financial support from the German Academic Scholarship Foundation and thanks the organizers and participants of the QuantHEP 2025 conference at Lawrence Berkeley National Lab for valuable discussions contributing to this work.
\end{acknowledgments}

\bibliographystyle{apsrev4-1}
\bibliography{MagicBarrier}

\newpage
\section{Supplemental Material}
\subsection{1. Entanglement spectra}

As explained in the main text, the spectrum of the entanglement Hamiltonian of the lattice gauge theory exhibits a coupling independent structure in the quantum ergodic regime. This was hinted at in Fig.~\ref{fig:fig4_coupling_dependence}~(c) for anti-flatness, constructed from the second and third moments of the spectrum. Here, we show that the coupling independence is also visible in the spectrum itself, by comparing the eigenvalues of the entanglement Hamiltonian $H_A=-\log\rho_A$ at different couplings but at the same stage during the thermalization process.

\begin{figure*}[h]
    \centering
    \includegraphics{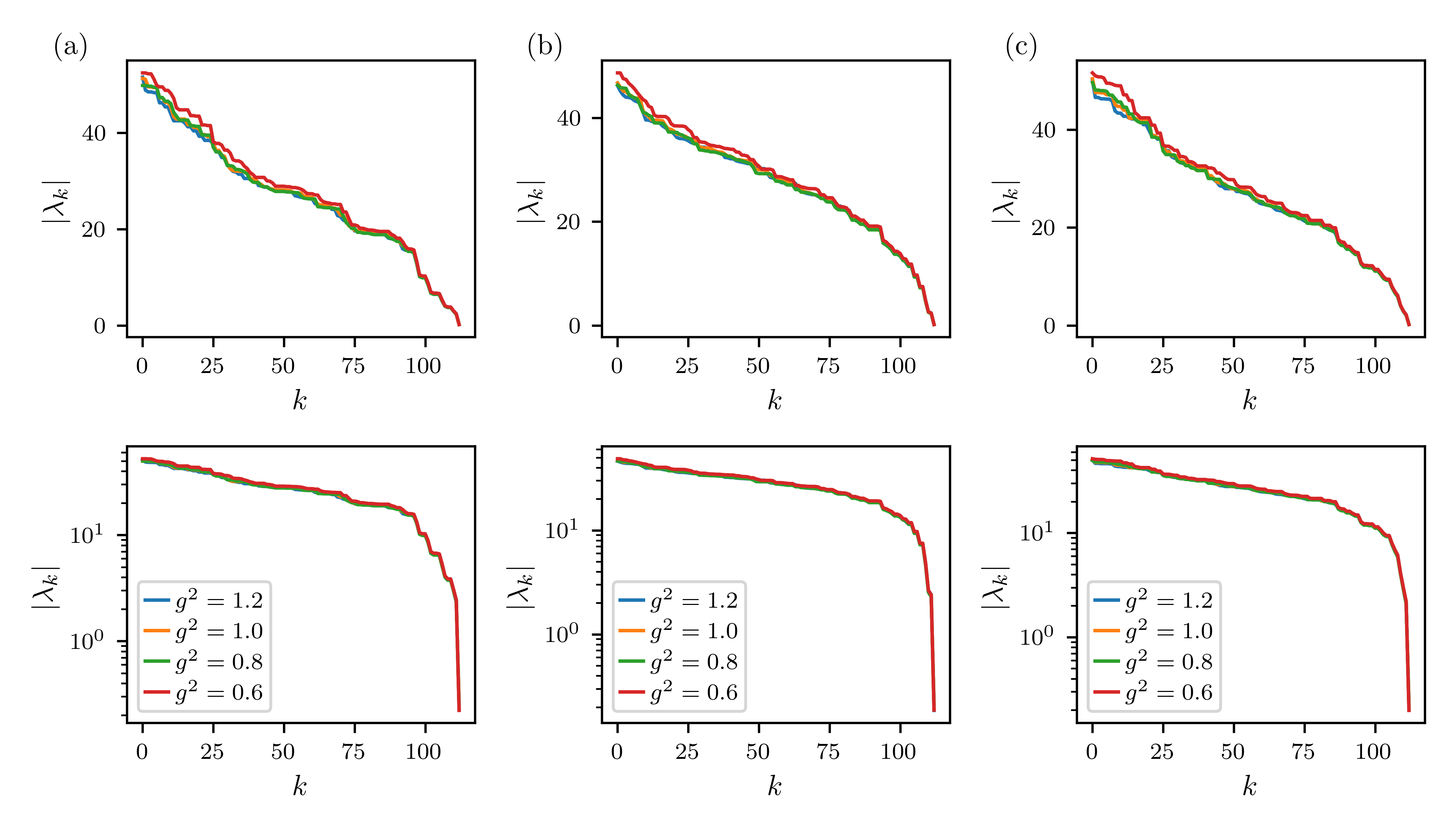}
    \caption{Entanglement spectra of three randomly chosen electric basis states at time $t_{\rm MB}$ for different couplings $g^2\in\{1.2, 1.0, 0.8, 0.6\}$ in the ergodic regime. The subsystem is a small one-plaquette piece with cut links in the middle of an aperiodic seven-plaquette chain with $\jmax=1$ and asymmetric boundary conditions $\{j_{\rm ext}\}=\{0,0,0,1\}$. The states lie in the highly excited energy window $E-E_0\in [26.48, 26.98]$ at coupling $g^2=0.6$. (a), (b), and (c) show the eigenvalues of the entanglement Hamiltonian $H_A=-\log\rho_A$ for the three states, respectively. Upper (lower) panels have a linear (logarithmic) scale.
    }
    \label{fig:figSUP_spectra_tMB}
\end{figure*}

Figure~\ref{fig:figSUP_spectra_tMB} shows the entanglement spectra for three randomly chosen electric basis states from the ensemble studied in Fig.~\ref{fig:fig4_coupling_dependence}, at couplings $g^2 \in \{1.2, 1.0, 0.8, 0.6\}$. For each coupling, the spectrum is evaluated at the corresponding coupling-dependent time $t_{\rm MB}$. In other words, we compare the least flat entanglement spectra of individual states during thermalization for various ergodic couplings.

As shown in the upper panels, the eigenvalues of the entanglement Hamiltonian exhibit a coupling independent spectral shape, with only minor deviations for large eigenvalues. Since large eigenvalues of $H_A$ correspond to exponentially small eigenvalues of $\rho_A$, we also display the spectra on a logarithmic scale in the lower panels.
From the perspective of the reduced density matrix, the spectra agree remarkably well across the different ergodic couplings. Deviations can be attributed to the uncertainty in determining $t_{\rm MB}$, which is about $\Delta t \approx 0.05$ in the time regime around $t \approx 0.3$.

\subsection{2. Mixed-field Ising model}

In the main text we speculate that the central result of this work -- a barrier-like maximum in the anti-flatness $\mathcal{F}_A$ of the entanglement spectrum during the time when local equilibration of the entanglement entropy $S_A$ is most rapid, which persists across the ergodic coupling regime -- generalizes to other quantum chaotic systems. To test this hypothesis, we investigate another well-studied chaotic quantum many-body system, the mixed-field Ising model \cite{Chiba:2024non}
\begin{align} \label{eq:mixed_Ising_Hamiltonian}
    H=J\sum_i\hat{\sigma}_i^z\hat{\sigma}_{i+1}^z + h_z\sum_i\hat{\sigma}_i^z + h_x\sum_i\hat{\sigma}_i^x \; ,
\end{align}
where $\hat{\sigma}_i^z$ and $\hat{\sigma}_i^x$ are the Pauli $Z$ and $X$ matrices of the spin at site $i$ and $h_z$, $h_x$ are the longitudinal and transverse field strengths, respectively. For $J$, $h_z$, $h_x \neq 0$ the system is proven to be non-integrable \cite{Chiba:2024non} and shows robust quantum chaotic behavior for the case that all three parameters have comparable magnitudes \cite{Kim:2013bal}. A well-studied parameter choice is $(J, \, h_z, \, h_x)=\left(1,\, (\sqrt5+5)/8, \, (\sqrt5+1)/4\right)\approx (1, 0.9045, 0.809)$, for which the system was also numerically shown to obey the strong eigenstate thermalization hypothesis \cite{Kim:2014eth}.

In comparison with the gauge theory in our main study, $h_z$ and $h_x$ take on the role of the coupling $g^2$. $h_x$, in particular, controls the strength of the off-diagonal matrix elements in the Hamiltonian and thus the scrambling rate of the computational basis states, which are the eigenstates of the diagonal $\hat{\sigma}^z$ terms. The range of transverse field couplings $h_x$ studied here is listed in Table \ref{tab:hx.dSdtmax}.

\begin{figure*}
    \includegraphics[]{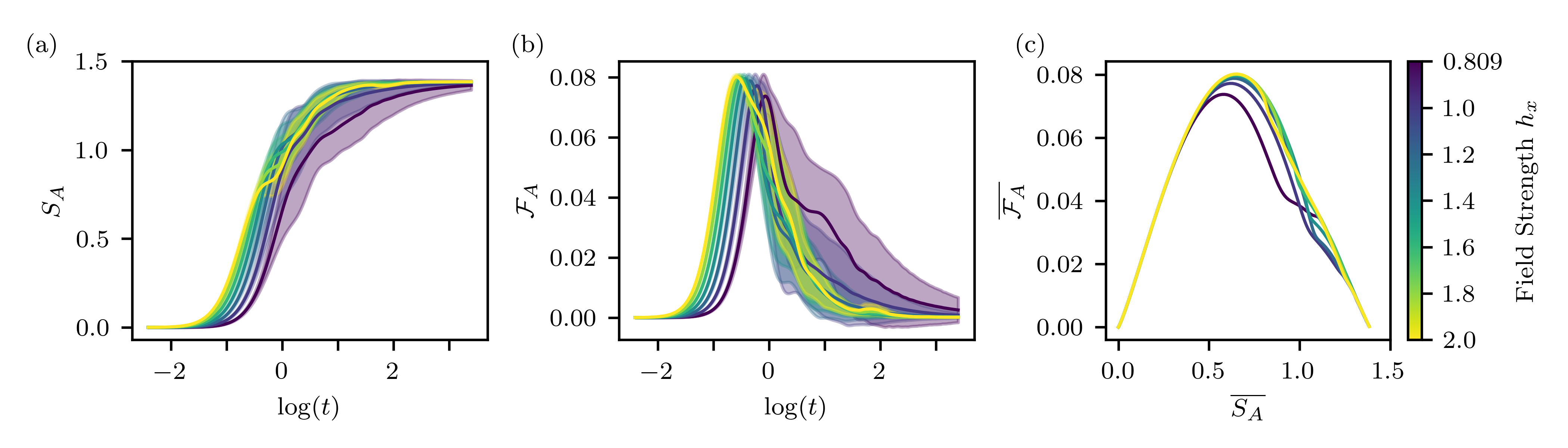}
    \caption{Coupling dependence of entanglement dynamics for a mixed-field Ising chain. Real-time evolution of (a) $S_A$ and (b) $\mathcal{F}_A$ for a subsystem of two spins (spins 7 and 8) in the middle of a chain of $N=15$ spins for the transverse couplings $h_x$ listed in Table \ref{tab:hx.dSdtmax} and fixed $J=1$ and $h_z = (\sqrt{5}+5)/8 \approx 0.9045$. The initial states are computational basis states chosen such that for $h_x=(\sqrt5 +1)/4 \approx 0.809$, the smallest coupling considered here, they lie in the highly excited energy window $E-E_0 \in [18.87, 19.17]$. This ensemble contains 2625 states. The solid lines and bands in (a) and (b) are the ensemble means and $1\sigma$-bands.
    (c) Relations between the ensemble means $\overline{S}_A$ and $\overline{\cF}_A$ for different $h_x$ couplings, which exhibit very mild coupling dependence up to times $t\sim \mathcal{O}(1)$, which corresponds to the entanglement entropies reaching approximately half of their maximum values. All chosen couplings exhibit the same qualitative behavior, with the only exception occurring for the smallest coupling $h_x=(\sqrt5 +1)/4 \approx 0.809$, which lies right at the edge of the ergodic regime, where the mean restricted gap ratio starts to deviate from the GOE prediction of $\langle r \rangle_{\rm GOE} \approx 0.5307$ (see Fig.~\ref{fig:fig7_mean_restricted_gap_ratio_Ising}). 
    }
    \label{fig:fig6_coupling_dependence_Ising}
\end{figure*}

Therefore, repeating the calculations of Fig.~\ref{fig:fig4_coupling_dependence} for the mixed-field Ising model with varying $h_x$ while fixing $J=1$ and $h_z=(\sqrt5+5)/8$ and taking an ensemble average over many highly excited states should show a qualitatively similar behavior. Indeed, in Fig.~\ref{fig:fig6_coupling_dependence_Ising} one can observe the same barrier-like maximum of the anti-flatness $\mathcal{F}_A$ that is highly correlated with the period of rapid growth of the entanglement entropy for a subsystem of two spins.

\begin{figure}[h]
    \centering \hspace*{-2em}
    \includegraphics[width=0.89\linewidth]{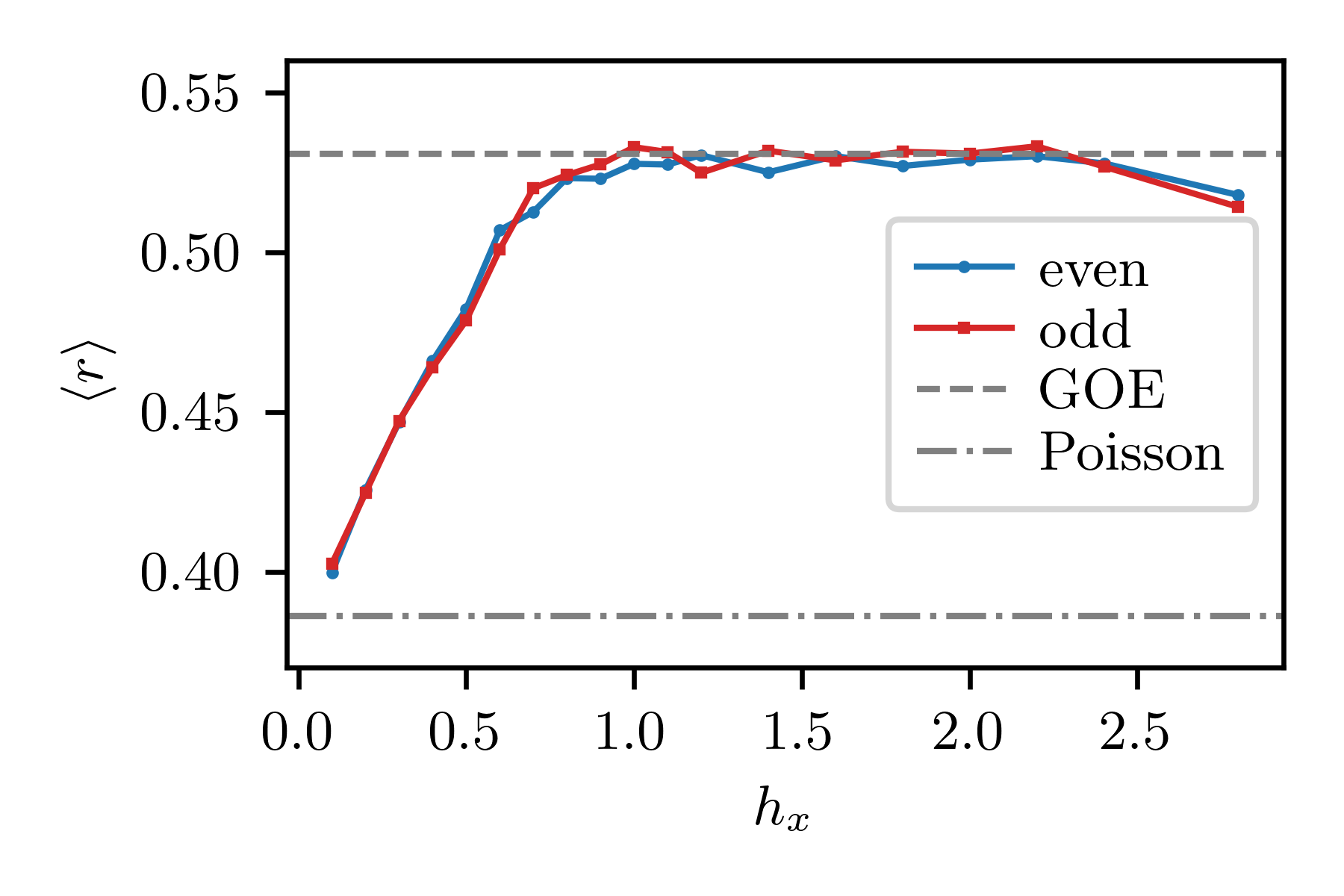}
    \caption{
    Coupling dependence of the mean restricted gap ratio $\langle r \rangle$ for the mixed-field Ising chain with $L=15$ spins. The parameters $J=1$ and $h_z=(\sqrt5+5)/8$ are fixed, while $h_x$ is varied. $\langle r \rangle$ is computed for the even (blue solid line) and odd (red solid line) parity sectors, which consist of 16512 and 16256 eigenstates, respectively. For both sectors, $\langle r \rangle$ is close to the GOE prediction (gray dashed line) of $\langle r \rangle_\mathrm{GOE} \approx 0.5307$ for $0.8 \lesssim h_x \lesssim 2.4$. In the integrable limit $h_x \to 0$ with only the longitudinal field, both sectors approach the Poisson prediction of $\langle r \rangle_\mathrm{P} \approx 0.3863$ (gray dash-dotted line) \cite{Atas:2013gap}. 
    }
    \label{fig:fig7_mean_restricted_gap_ratio_Ising}
\end{figure}

The system in Fig.~\ref{fig:fig6_coupling_dependence_Ising} contains $L=15$ spins. Its Hilbert space dimension of 32758 is comparable in magnitude to that of the seven-plaquette SU(2) chain with open boundary conditions explored in Fig.~\ref{fig:fig4_coupling_dependence}. Moreover, increasing the coupling $h_x$ inside the ergodic regime ($0.8 \lesssim h_x \lesssim 2.4$, see Fig.~\ref{fig:fig7_mean_restricted_gap_ratio_Ising}) leads, as expected, to an earlier onset of entanglement and anti-flatness growth due to faster scrambling of the computational basis states under unitary time evolution. As in the gauge theory case, this coupling dependence almost disappears if the anti-flatness is plotted against the entanglement entropy tracking thermalization progress instead of time. The curves are almost identical up to times $t\sim \mathcal{O}(1)$, where the entanglement entropy reaches roughly half of its maximal value. The only coupling for which the curves slightly deviate from the rest is $h_x=(\sqrt5+1)/4\approx 0.809$, which lies at the lower edge of the ergodic regime and has a mean restricted gap ratio slightly below the GOE value (see Fig.~\ref{fig:fig7_mean_restricted_gap_ratio_Ising}).

\begin{table}[htb]
\centering
\begin{tabular}{|c|c|c|}
\hline
$h_x$ & $(d\overline{S}_A/dt)_{\mathrm{max}}$ & $(\overline{\mathcal F}_A)_{\mathrm{max}}$ \\
\hline
$(\sqrt{5}+1)/4$ & 1.332 & 0.0737 \\
1.0 & 1.719 & 0.0772 \\
1.2 & 2.006 & 0.0788 \\
1.4 & 2.186 & 0.0796 \\
1.6 & 2.284 & 0.0799 \\
1.8 & 2.327 & 0.0801 \\
2.0 & 2.334 & 0.0802 \\
\hline
\end{tabular}
\caption{Transverse field strengths $h_x$ explored here, associated maximal growth rate of the entanglement entropy $(d\overline{S}_A/dt)_{\mathrm{max}}$, and the peak anti-flatness $(\overline{\mathcal F}_A)_{\mathrm{max}}$.}
\label{tab:hx.dSdtmax}
\end{table}

Around $t\sim \mathcal{O}(1)$, both the mixed-field Ising model (see Fig.~\ref{fig:fig6_coupling_dependence_Ising}) and the SU(2) plaquette chain (see Fig.~\ref{fig:fig4_coupling_dependence}) exhibit small oscillatory structures in the time evolution of the entanglement entropy and the anti-flatness. We attribute these features to finite-size and microscopic effects leading to low-frequency modes that become relevant at times comparable to the local interaction scale, before self-averaging and coarse-grained thermalization set in.

We observe a similar increase in the maximal growth rate of the entanglement entropy $(d\overline{S}_A/dt)_{\mathrm{max}}$ with the transverse field strength $h_x$, which reflects the level of integrability breaking, as for $1/g^2$ in the SU(2) system (see Table \ref{tab:hx.dSdtmax}).  The growth rate scales approximately linearly with $h_x$, analogously to the coupling dependence of the magnetic energy in Eq.~(\ref{eq:KS_Hamiltonian}). For $h_x \gtrsim J$ the growth rate $(d \overline{S}_A / dt)_\mathrm{max}$ saturates. We attribute this behavior to the fact that, in this regime, the transverse field predominantly enhances local spin precession, while the generation of nonlocal correlations -- and hence entanglement between different sites -- remains constrained by the interaction term proportional to $J$. As a result, the entanglement growth rate becomes limited by the interaction strength $J$ rather than the magnitude of the transverse field.

Taken together, these results for the mixed-field Ising model suggest that the barrier-like peak in the anti-flatness, closely correlated with rapid entanglement growth, is not specific to the SU(2) plaquette system. Rather, this behavior likely generalizes to a broad class of quantum chaotic spin chains, providing support for the conjecture that the emergence of an anti-flatness barrier is a generic feature of thermalization dynamics in chaotic quantum many-body systems.


\end{document}